\documentclass[epsf,preprint,aps]{revtex4}
\usepackage{latexsym}
\usepackage{amsfonts}
\usepackage{graphicx}
\usepackage{epsfig}
\usepackage{amsmath, amsthm, amssymb}

\begin{document}
\title{Effect of pressure on the kinetics of bulge bases in small RNAs}

\author{Pradeep Kumar, J. Lehmann, A. Libchaber}

\affiliation{ Center for Studies in Physics and Biology, The
  Rockefeller University, 1230 York Avenue, New York, NY 10021 USA
  \\ } (kumar2010d.tex 22 July 2010)

\begin{abstract}

Using molecular dynamics simulations, we study the effect of pressure
on the binding propensity of small RNAs by calculating the free energy
barrier corresponding to the looped out conformations of unmatched
base, which presumably acts as the binding sites for ligands. We find
that the free energy associated with base looping out increases
monotonically as the pressure is increased. Furthermore, we calculate
the mean first passage time of conformational looping out of the base
bulge using the diffusion of reaction coordinate associated with the
base flipping on the underlying free energy surface. We find that the
mean first passage time associated with bulge looping out increases
slowly upon increasing pressures $P$ upto $2$~kbar but changes
dramatically for $P>2$~kbar. Finally, we discuss our results in the
light of the role of hydration shell of water around RNA.

\end{abstract}

\maketitle

\noindent
\section{Introduction}
\bigskip

RNA molecules are very diverse both structurally and
functionally~\cite{Cech1987}. Apart from having the regular helical
purine-pyramidene base pairs, RNA molecules are also found to have
many other secondary structures (motifs) such as loops, knots and
bulges etc~\cite{Wyatt1989,Hermann2000}. The presence of such
structural motifs is found to play a role in binding of different
molecules to RNA~\cite{Lustig1998}. For many protein binding RNAs, it
was found that the frequency of adenosine bulge at the binding site is
very high. The presence of a bulge may change the conformational
flexibility of an RNA~\cite{Patel1982,Woodson1989} and hence more
internal surface area of RNA is available for any chemistry. Moreover,
the presence of the bulges does not only change the conformational
flexibility but the bulges themselves may just flip out exposing the
internal regions of a RNA to solvent and ligands. It has been shown
that the bulge base looping out is highly sensitive to the bulge bases
and their neighbors~\cite{Patel1982,Woodson1989}, which makes the
question of generality of any picture of base bulge looping out
difficult.

Recent state of the art computer simulations of small RNAs have shed
light on the base bulge looping out
processes~\cite{Andre06,Feig2001,Auffinger2007} in few of the possible
cases. Specifically, these works have studied the free energy barriers
associated with bulge base looping out process~\cite{Andre06}. For
example A. Barthel and M. Zacharias studied single uridine and
adenosine bulge structures and their looping out using the torsional
angle which measures the degree of looping out of bulge bases from the
local helical plane. They find that the conformational free energy
change in the case of adenosine bulge in a complete looping out
process is higher by $1.5$~kcal.mol$^{-1}$ as compared to the
adenosine bulge~\cite{Andre06}, suggesting that in a base nonspecific
binding process a structure with single uracil bulge base would have
higher propensity to flip out of helical plane. Although, a wealth of
literature is available on the base looping from the helical plane at
ambient conditions; the changes in the kinetics of bulge base flipping
is rather unexplored at conditions away from ambient conditions. 

It is known that water's hydrogen bond network and so the local
structure of liquid water changes upon changing thermodynamic
conditions, giving rise to anomalous changes in the dynamics and
thermodynamics of water and aqueous systems. The structural stability
and kinetics of proteins (where structural stability usually implies
kinetically functional) as a function of pressure and temperature. The
solvation barrier in the case of proteins plays an important role both
in the hydrophobic collapse of the polypeptides as well as the
stability of these structures as a function of pressure and
temperature. In this paper, we will investigate another class of
molecules which are involved in ligand binding kientics and are very
stable at high pressures and temperatures. One can ask questions
whether the structural stability in these cases will also imply
kinetic functionality. We specifically study the binding of a purine A
based molecules to small RNAs with a single A-bulge structural
motif. This study is relevant to understand the self-aminoacylation of
small ribozymes~\cite{illanga,Lehmann2007} in different thermodynamic conditions. It is widely
believed that for the purine A-based molecules such as ATP to bind to
such a bulge, the A-bulge base has to flip out, which allows the
purine A-based molecules to come and stack into the bulge
configuration. In other words, for the binding to proceed the A-bulge
has to overcome both the bending regidity and solvation energy barrier
to solvate in water. The solvation of different substances in water is
a widely studied
problem~\cite{tanford,kdill,lum,ashbaugh,garde,widom}. Indeed, studies
of apolar solutes in water shows an ellipitic region in the
pressure-temperature plane in which water behaves as a bad solvent and
hence less solubility of these subtances~\cite{buldyrev07}. However,
study of solvation barrier of base flipping in case of RNA/DNA is not
studied.

In this paper, we study the effect of pressure on the kinetics of
bulge base looping out of a double strand RNA with single adenosine
bulge. In section II, we discuss the method, in section III we present
the results for the free energy barrier associated with torsional
deviation of the bulge base from the local helical backbone, in
section IV we present a mean first passage time calculation associated
with looping out process and finally we discuss and summarize the
results.

\begin{figure}[htb]
\begin{center}
\includegraphics[width=14cm]{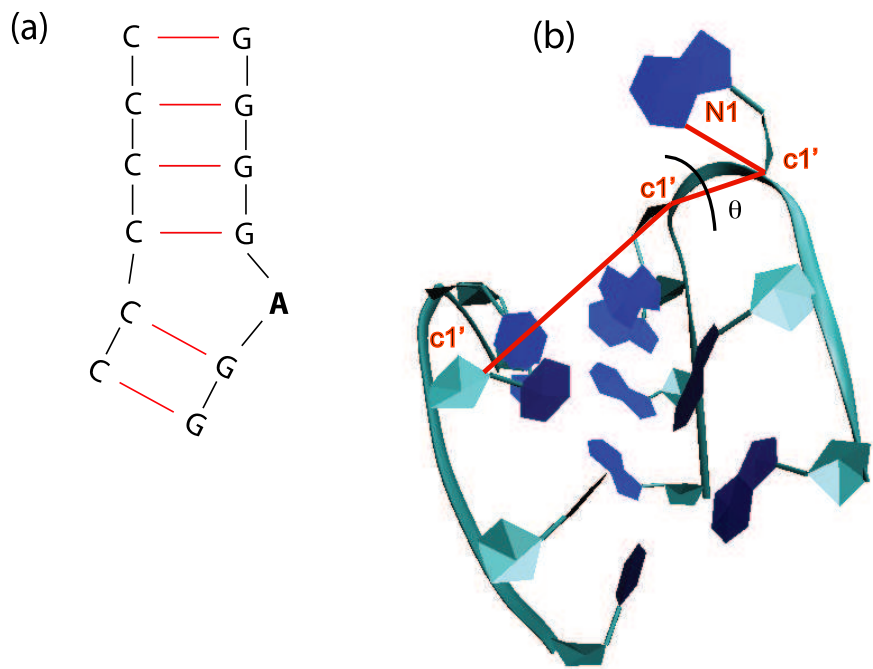}
\end{center}
\label{fig:figbulge}
\caption{(a) A 2-dimensional schematics of the double strand RNA
  structure used in this study with a single {\it A} bulge in
  5'-GGGGAGG-3'5'-CCCCCC-3'. (b) The three-dimensional structure and
  the Definition of the torsional angle (C1'C1'C1'N1) chosen as the
  reaction coordinate for calculation of the free energy.}
\label{fig:fig2}
\end{figure}

\bigskip
\section{System and Method}
\bigskip
The energy minimized starting structure of the RNA with single {\it A}
(5'-GGGGAGG-3'/5'-CCCCCC-3') bulge was created using the NAB/Nucgen
module of the Amber10 program suite (Fig.~\ref{fig:figbulge}). The
RNA structure was then solvated in ~4000 TIP3P water molecules such
that there was about $1.0$~nm space left between the boundary of the
box and the RNA solute atoms. For electro-neutrality $14$ Na+ counter
ions were added to the solution. Energy minimizations were carried
using the steepest descent ($1000$ steps) in
GROMACS3.3.3~\cite{gromacs1,gromacs2} with keeping the RNA atoms
fixed. Simulations were carried out using the GROMACS3.3.3 program
with the Amber99 force field with a periodic boundary conditions and
integration time step of $2$~fs. The long range electrostatic
interactions were treated with the particle-mesh-ewald (PME) method.
After minimization the system was slowly heated to temperature
T=$300$~K with positional restraints.

After the position restrained simulations, unrestrained Molecular
Dynamics (MD) was carried out at four different values of the
pressures $P=1,1000,2000,3000$~atm. Thermal equilibrium at a constant
temperature $T=300$~K and different pressures was achieved using
Berendsen thermostat and barostat respectively. The final equilibrated
conformation was then used for as the starting conformation for
umbrella sampling at different pressures.

To quantify the relative propensity of binding of RNA at different
pressures, we chose the dihedral angle $\theta$ (C1'C1'C1'N1) (see
Fig.~\ref{fig:fig2}) as the reaction coordinate for calculation of
free energy. Since the conformational changes are very slow and an
equilibilirium sampling of torsional angles require much larger time
scales than computationally feasible, we use biasing potential to
calculate the free energy. Umbrella sampling method was used to
calculate the relative free energy of bulge base looping out
conformations for the RNA shown in Fig.~\ref{fig:fig2}. Harmonic
umbrella biasing potentials $U_{bias,\theta_i} =
k(\theta-\theta_{i})^2$ with a force constant $k=0.05
kcal.mol^{-1}.deg^{-2}$ were distributed uniformly along the reaction
coordinate $\theta$ at an interval $\Delta
\theta_{ref}=5^{o}$. Consecutive sampling windows of $\theta$ were
started from equilibrium structure of last run. For each values of the
umbrella sampling window, we run a $2$~ns simulation and record the
value of $\theta$ every $0.2$~ps. The final potential of mean force
(PMF) was calculated using the WHAM (weighted histogram
method)~\cite{wham}. The unbiased probability distribution $P(\theta)$
at a given temperature $T$ under WHAM is given by

\begin{equation}
P(\theta) =
\frac{\sum_{i=1}^{N}n_i(\theta)}{\sum_{i=1}^{N_{sim}}n_{i}e^{[F_i-U_{bias,i}(\theta)]/k_{B}T}}
\label{eq:eq1}
\end{equation}
where $N_{sim}$ is the number of sampling window (simulations), $n_i$
is the number of counts in the bin associated with $\theta$,
$U_{bias,i}$ is the biasing potential, and $F_{i}$ free energy from
simulation $i$ and is given by
\begin{equation}
F_{i} = -k_{B}Tln[\sum_{\theta_{bins}}P(\theta)e^{(-U_{bias,i}(\theta)/k_BT)}]
\label{eq:eq2}
\end{equation}
where $\theta_{bins}$ is the number of bins for the individual
sampling window. The equations ~\ref{eq:eq1} and ~\ref{eq:eq2} are
iterated to obtain the self consistent value of $P(\theta)$. The value
of $P(\theta)$ depends on the time scale of simulations and hence long
simulations are needed for a good convergence of the free energy.

\begin{figure}
\begin{center}
\includegraphics[width=12cm]{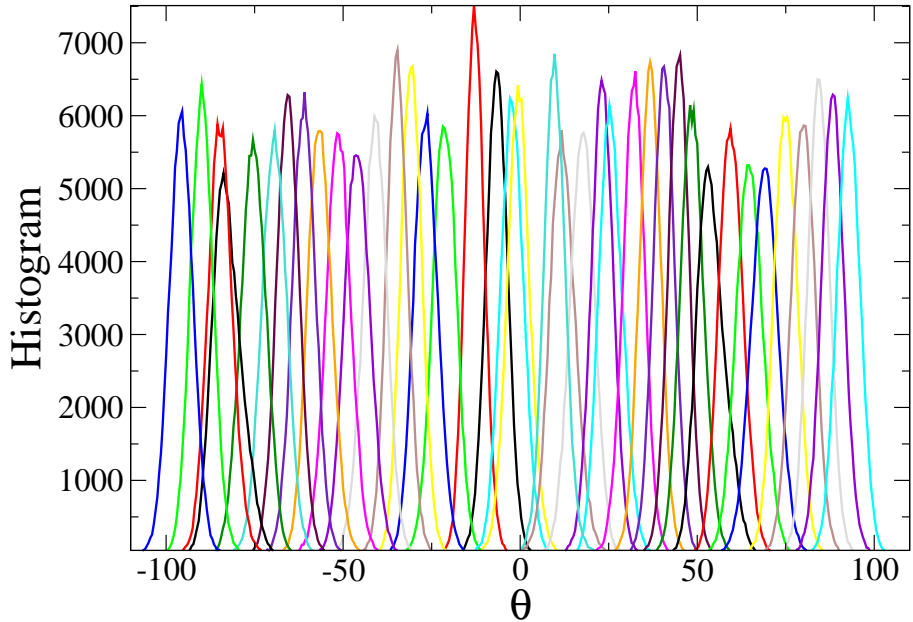}
\end{center}
\caption{A typical histogram for biasing potential centered at
  different values of $\theta$.}
\label{fig:hist}
\end{figure}

In Fig.~\ref{fig:hist}, we show typical histograms along the reaction
coordinate $\theta$ for biasing potential centered at different values
of $\theta$ with $\Delta \theta=5^{o}$. We obtain smooth histograms
for all the sampling windows for all the trajectories.

\section{Kinetics of single {\it A} bulge at ambient and elevated pressures}

\begin{figure}[htb]
\begin{center}
\includegraphics[width=12cm]{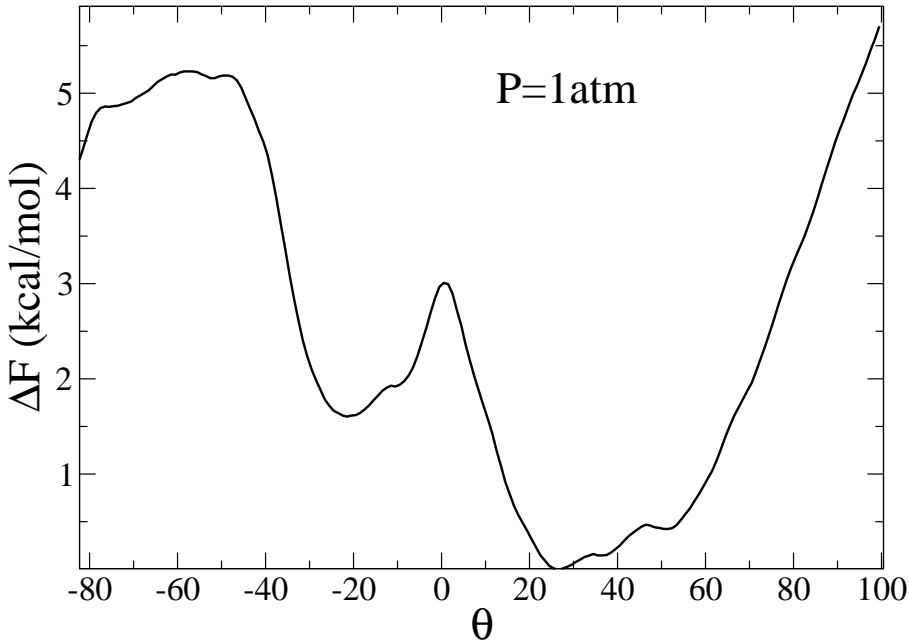}
\end{center}
\caption{Free energy as a function of the torsional angle $\theta$ for
  single A bulge at $P=1$~atm. The free energy profile associated with bulge flipping has
  two minima as found in earlier studies~\cite{Andre06}, separated by
  an energy barrier of $~3.0$~kcal/mol.}
\label{fig:freeEA}
\end{figure}

In Fig.~\ref{fig:freeEA}, we show $F(\theta)$ as a function of
$\theta$ for the RNA structure with $A$-bulge at $P=1$~atm. Negative
value of $\theta$ corresponds to deviation towards minor groove while
the positive values corresponds to deviation towards major groove. We
find that $F(\theta)$ has characteristic two minima centered around
$\theta\approx-20^{o}$ and $\theta\approx30^{o}$ as reported in
earlier studies of single A bulge~\cite{Andre06}. Note that the
definition of the torsional angle $\theta$ is different from the one
used Ref.~\cite{Andre06} and hence different values of $\theta$. As we
can see from figure~\ref{fig:freeEA}, the orientation of the A-bulge
at more stable minimum is tilted slightly along the major groove while
the second minimum at $\theta\approx-20^{o}$ is presumably due to the
base triplet formation with the neighboring bases. The free energy
difference between these two minima is $\approx 3$kcal/mol. suggesting
that although $\theta=30^{o}$ is relatively a more stable minimum
configuration, the thermal fluctuations at $T=300$~K
($\approx0.60$~kcal/mol) is sufficient enough for the bulge to get
displaced of the free energy minimum configuration. Due to a large
free energy barrier ($\Delta F\approx 5kcal/mol)$ a complete looped
out conformation of single A-bulge is less favorable and hence
consistent with experimental fact that a single {\it A} does not bind
efficiently with the incoming ligands.

\begin{figure}[htb]
\begin{center}
\includegraphics[width=12cm]{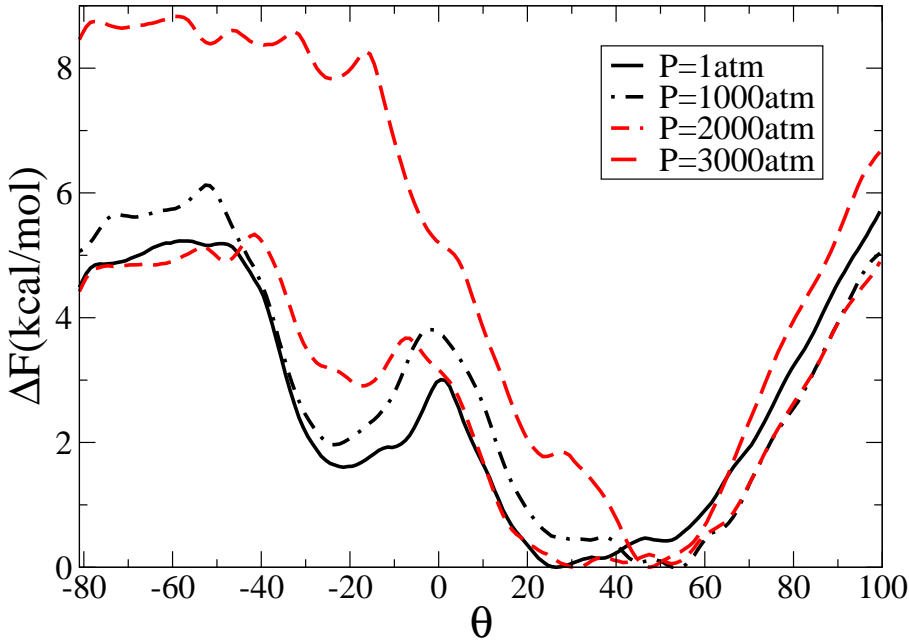}
\end{center}
\caption{Free energy as a function of the torsional angle $\theta$ for
  different pressures. The free energy barrier between the two minima
  slowly disappears at pressure $P > 2000$; suggesting that the
  transient barrier that is produced at atmospheric pressures due to
  the partial base triplet hydrogen bonding of the bulge base with the
  neighboring bases is almost broken at higher pressures.}

\label{fig:freeEAallp}
\end{figure}

Figure~\ref{fig:freeEAallp} shows $F(\theta)$ as a function of
$\theta$ for different $P$. We find that for pressures up to
$2000$~atm $F(\theta)$ has the characteristic two minima as we find in
the case of $P=1$~atm. However, as the pressure in increased, the
minimum at $\theta\approx-20^{o}$ becomes shallower and disappears for
$P>2000$~atm, suggesting that at $P>2000$~atm the base triplet
formation of the bulge base with the neighboring bases does not occur
during the looping process. Moreover, the free energy barrier between
the two minima changes just a little upon increasing pressure for
$P<2000$~atm. For $P=3000$~atm, free energy barrier for the flipped
out state changes drastically where $\Delta F\approx8.7$~kcal/mol,
suggesting that the propensity of single A-bulge base looping out from
the local helical plane would decrease upon increasing pressure and so
the binding propensity of of incoming ligands. Furthermore, we note
that although the free energy barrier associated with complete looped
out conformations is finite is for all the pressures the final looped
out state is not a free energy minimum state hence a RNA structure
with single A bulge would not favor ligand binding. To verify this we
ran $20$~ns long simulations to see if there is any bulge looping out
event at different pressures and we do not observe any.

In order to calculate the effective rate $k_{eff}$ of transition
from a stacked to looped out conformation, we use Langevin
equation~\cite{KampenBook,Gardiner}. Assuming the diffusion of the
reaction coordinate $\theta$ on an underlying free energy surface the
dynamics of $\theta$ is governed by
\begin{figure}[htb]
\begin{center}
\includegraphics[width=10cm]{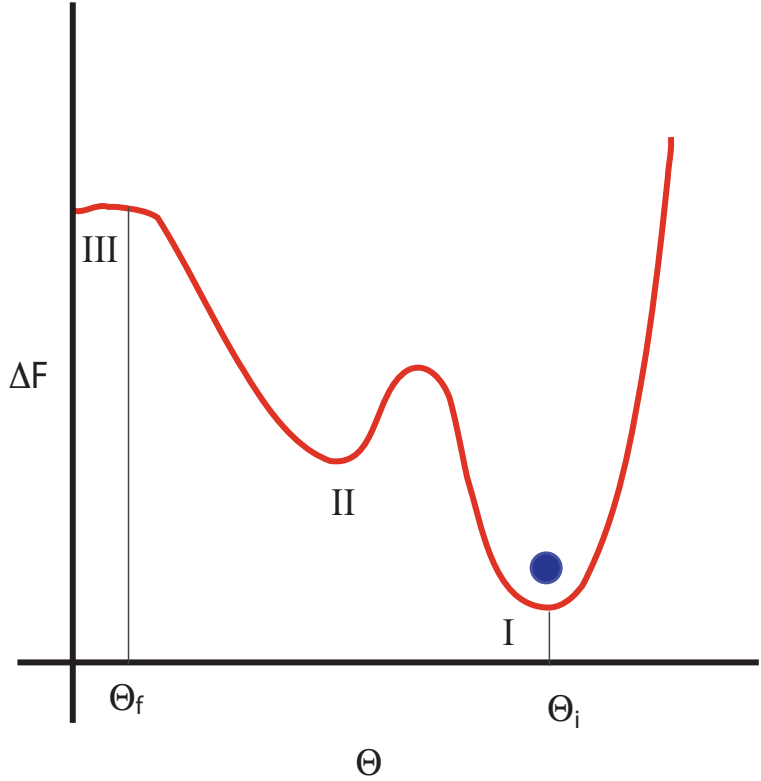}
\end{center}
\caption{Schematic of free energy as a function of the torsional angle
  $\theta$ used for the calculation of the effective time scale of
  bulge base flipping kinetics.}
\label{fig:freeschematic}
\end{figure}
\begin{equation}
\dot{\theta} = -\frac{D}{k_BT}\partial_{\theta}F(\theta) + f(t)
\end{equation}
where $\theta$ is the torsional deviation and $D$ is the diffusion
constant and $f(t)$ is the thermal noise with zero mean, $\langle
f(t)\rangle =0$ and delta function correlation, $\langle
f(t)f(0)\rangle=2D\delta(t)$. In the high friction limit, the
probability $\rho(\theta,t)$ of finding the system with reaction
coordinate $\theta$ after time $t$ is given by the Smoluchowsky
equation:
\begin{equation}
\rho(\theta,t) = L\rho(\theta,t)
\end{equation}
where $L$ is the Fokker-Planck operator given by $L=\partial_\theta
e^{-\beta F(\theta)}D\partial_\theta e^{\beta F(\theta)}$ and
$\beta=1/k_BT$. The mean first passage time $t(\theta_{i})$ associated
with crossing the barrier from any coordinate $\theta_{i}$ to final
state $\theta_{f}$ is given by (see Fig.~\ref{fig:freeschematic})
\begin{equation}
t(\theta_{i}) = \int_{\theta_{i}}^{\theta_{f}}dy\frac{1}{D}\int_{\theta_r}^{y}dx ~e^{(\beta(F(y)-F(x)))},
\label{eq:t}
\end{equation}
where $\theta_r$ and $\theta_s$ denote the reflecting and absorbing
boundaries respectively. We choose $\theta_{f}=80^{o}$ as the final
looped out state and initial state is chosen to the values of $\theta$
where the free energy curve has the deepest minimum for respective
pressures. The effective rate $k_{\rm eff}$ of transition from $I$ to
$III$ would then be given by $\frac{1}{t(\theta_i)}$. The reflecting
boundary was chosen to be at $\theta=-100^{o}$ where the relative free
energy is $\approx 10 k_BT$. Using Eq.~\ref{eq:t}, we calculate the
value of $t(\theta_i)$ for different pressures. We list the values of
$t(\theta_i)$ in table~\ref{table:tab3} where $D$ is measured in
$deg^2/sec$. We find that $t(\theta_i)$ increases upon increasing
pressure within the error bars and increases sharply for $P=3000$~atm.

\begin{center}
\begin{table}
\begin{tabular}{|l|l|}

\hline \multicolumn{2}{|c|}{} \\ \hline
Pressure & $t(\theta_{i})D^{-1}10^6 (deg^2)$ \\ \hline

1 atm &  2.1625\\ \hline 

1000 atm &  9.8104 \\ \hline 

2000 atm &  5.3807 \\ \hline 

3000 atm &  1018 \\ \hline

\end{tabular}
\caption{Mean first passage times}
\label{table:tab3}
\end{table}
\end{center}

\section{Hydration shell and base flipping of RNA}

\bigskip

\begin{figure}[htb]
\begin{center}
\includegraphics[width=10cm]{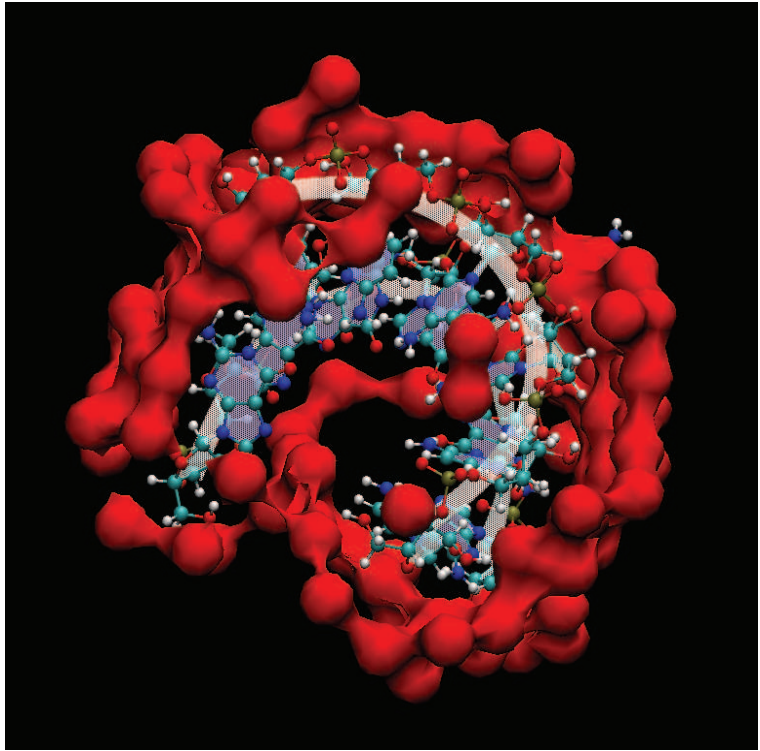}
\end{center}
\caption{A typical hydration sheath around RNA.Water molecule is
  represented by a surface plots and the hydration shell is obtained
  as mentioned in section IV.}
\label{fig:hydration}
\end{figure}

As we have seen the sections above, that the base flipping kinetics
changes as the pressure is increased -- namely, the free energy
barrier for the bulge base to flip out increases with
pressure. Moreover, the transient barrier which presumably is due to
the base triplet formation of the bulge base with the neighboring
bases disappears at pressures $P > 2$~kbar. We note that $2$~kbar is
the pressure where most of the anomalies of liqid water disappears and
also the pressure at which hydrophobic barriers for small molecules in
water tend to vanish. Motivated by this we looked at the structure of
the solvation shell (first hydration shell) of water around RNA for
different pressures. We show a typical hydration shell around RNA in
Fig.~\ref{fig:hydration}. The hydration shell is calculated by finding
all the water molecules within a distance $R_C$ of RNA molecule. We
choose $R_C=0.223$~nm as the first minimum in the radial distribution
function of RNA and oxygen of water molecules (not shown here). We
find that $R_C$ is independent of the pressure.

\begin{figure}
\begin{center}
\includegraphics[width=10cm]{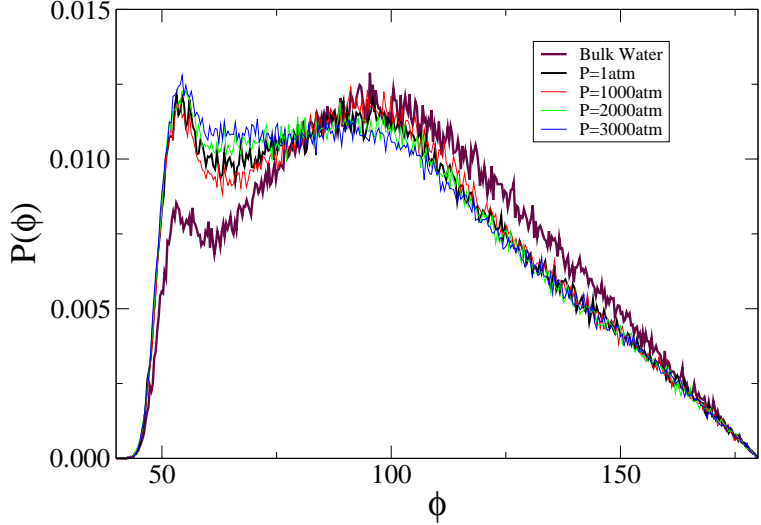}
\end{center}
\caption{Probability distribution function $P(\phi)$ of the OOO-angle
  $\phi$ of the hydration shell for various pressures. Note that a
  comparison with $P(\phi)$ for bulk water at $P=1$~atm suggests, that
  hydration shell around RNA is more disordered and this disorder
  increases monotonically upon increasing pressure.}
\label{fig:distangle}
\end{figure}

We first looked at whether the observed change in the pressure
dependence of the kinetics is a result of ordering of water around
RNA. To this effect, we calculated the OOO-bond angle $\phi$ and its
distribution $P(\phi)$ of a central water molecule and its nearest
neighbors in the hydration shell. If the water orders then $\phi$,
would be very close to the tetrahedral angle $109.47^o$. In
Fig.~\ref{fig:distangle}, we show $P(\phi)$ for pressures
$P=1,1000,2000,3000$~atm. For a comparison we also plot $P(\phi)$ for
bulk TIP3P water at atmospheric pressures. We find that, the second
peak corresponding to more ordered water of the distribution $P(\phi)$
shifts to smaller values of $\phi$, suggesting that the water
monotonically disorders upon increasing pressure. Moreover, we do not
find any significant sharp changes in $P(\phi)$ which could be
associated with sharp change in the free energy barrier observed at
$P=3000$~atm.

To quantify the ordering/disordering of water molecules around $RNA$,
we use the tetrahedral order
parameter$Q$~\cite{Chou98,errington,kumarPNAS09}. Tetrahedral order
parameter $Q$ quantifies how close a given water molecule and
its first shell neighbors form a structure close to a tetrahedron. In
general, $Q_k$ of $k^{th}$ molecules is defined as
\begin{equation}
Q_k = 1- \frac{3}{8}\sum_{i>j}\sum_{j}(cos\phi_{ikj}+1/3)^2
\end{equation}
where, the indices i, and j run over all four neighboring molecules
and $\phi_{ikj}$ the OOO-angle formed between the oxygens of central
molecule k and neighbors i and j.

Since, we only consider a thin hydration shell, the expression for
ensemble average $<Q>$ can be written as 
\begin{equation}
<Q> = 1 -
\frac{9}{4}\int_0^{\pi}(cos\phi_{OOO}+1/3)^2P(\phi_{OOO})d\phi_{OOO}
\end{equation}
In table~\ref{table:Q}, we list the average tetrahedral order
parameter $<Q>$ of the hydration shell for different pressures. For a
comparison, we also compute $<Q>$ for bulk water. Table~\ref{table:Q}
lists values of $<Q>$ for different pressures. We find that, $<Q>$
monotonically decreases upon increasing pressure and no sudden change
in $<Q>$ is seen at $P=3000$~atm.

\begin{center}
\begin{table}
\begin{tabular}{|l|l|l|}
\hline
\multicolumn{3}{|c|}{Average Tetrahedral Order Parameter $<Q>$} \\
\hline
Pressure & Bulk Water & Hydration Shell \\ \hline
1 atm & 0.467  & 0.396 \\ \hline
1000atm &  & 0.375 \\ \hline
2000 atm &   & 0.361 \\ \hline
3000atm &  & 0.348 \\ 
\hline
\end{tabular}
\caption{Average tetrahedral order parameter $Q$ of the first
  hydration shell of water around RNA at different pressures}
\label{table:Q}
\end{table}
\end{center}

\begin{figure}[htb]
\begin{center}
\includegraphics[width=16cm,height=8cm]{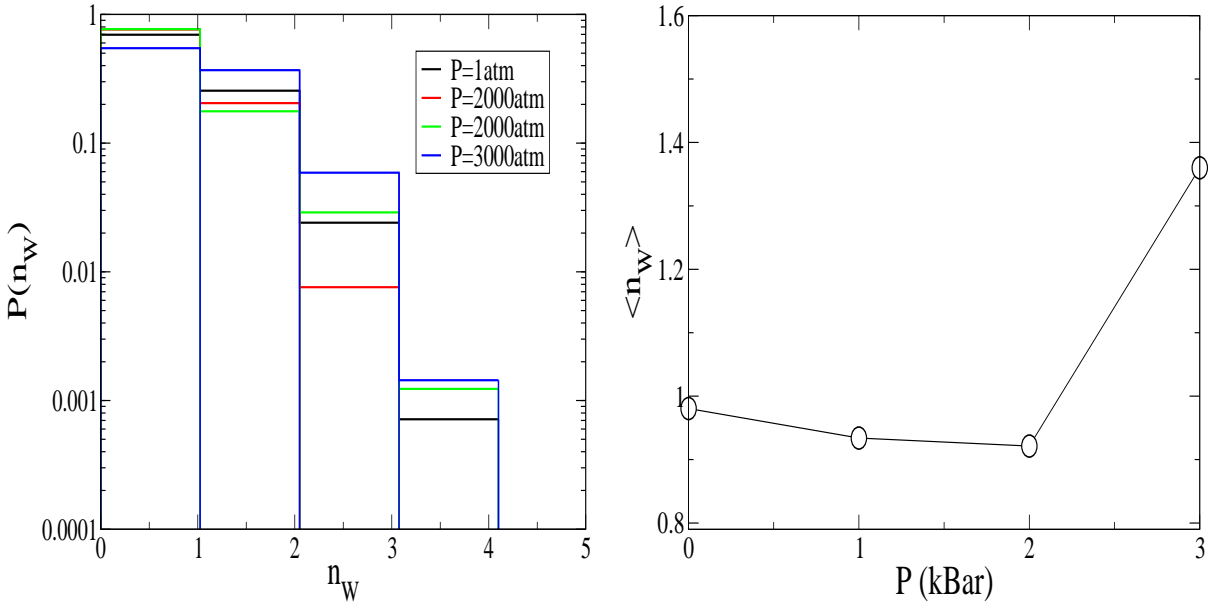}
\end{center}
\caption{(a) Probability distribution $P(n_W)$ of water molecules in
  the first hydration shell around the bulge base ``A'' for different
  pressures. Note that the probability of $3$ or $4$ water molecules
  increases as the pressure is increased. (b) Ensemble averaged
  $<n_W>$ as a function of pressure.}
\label{fig:nwater}
\end{figure}

Since, we did not see any sudden change in the ordering of hydration
shell around RNA that might lead to the base flipping barrier observed
at $P=3000$~atm, we next studied the hydration shell of the bulge base
and suprisingly, we find that at high pressures, the average number of
water molecules in the first hydration shell of the bulge base
increases, from an average of about $1.0$ to $~1.40$ (see
Fig.~\ref{fig:nwater}(b)). Moreover, we find that the distribution of
water molecules around the bulge base shows significant probability of
finding $3-4$ water molecules. To this end, we suggest that the
increased barrier of base flipping and the disappearance of the
transient barrier in the free energy barrier is due to the presence of
increased water molecules in the solvation shell around the bulge
base. The presence of more water molecules creates a competition
between the base triplet formation and the hydrogen bond formation
with the water molecules, which might have penetrated from the major
groove, and hence stabilizing the stacking of the bulge base.

\section{Summary and Discussions}

In summary, in this paper we have investigated the effect of pressure
on the kinetics of base flipping by calculating the free energy
barrier associated with looping out torsional angle of an adenosine
bulge base in a short double strand RNA. We find that, upon increasing
the pressure, the propensity or likelihood of base flipping
decreases. At pressure $P>2000$~atm, we see a sharp increase in the
free energy barrier. Further, we calculate the time scale of flipping
by mapping the problem of base flipping to a diffusion of reaction
coordinate on an underlying free energy landscape from which we
calculate the time scale of looping out of bulge base. We find that
the time scale increases upon increasing pressure and changes
dramatically at $P>2000$~atm. We associate this behavior with the
competing hydrogen bonds between the neighboring bases and increased
solvation of the bulge base at high pressures.

\bigskip

\acknowledgments

Authors would like to thank NSF grant no. PHY-0848815 for support. PK
acknowledges support from National Academies Keck Foundation Future
Initiatives Award.

\end{document}